\documentclass[draft]{agujournal2019}
\usepackage{url} 
\usepackage{lineno}
\usepackage[inline]{trackchanges} 
\usepackage{soul}
\usepackage{amssymb}
\usepackage[mathscr]{euscript}
\usepackage{appendix}

\linenumbers

\draftfalse

\begin{document}
\nolinenumbers


\title{Learning Generative Models for Lumped Rainfall-Runoff Modeling}

%

\authors{Yang Yang\affil{1}, Ting Fong May Chui\affil{1}}

\affiliation{1}{Department of Civil Engineering, The University of Hong Kong, Hong Kong SAR, China}

\correspondingauthor{Ting Fong May Chui}{maychui@hku.hk}

%


\begin{abstract}

This study presents a novel generative modeling approach to rainfall-runoff modeling, focusing on the synthesis of realistic daily catchment runoff time series in response to catchment-averaged climate forcing. Unlike traditional process-based lumped hydrologic models that depend on predefined sets of variables describing catchment physical properties, our approach uses a small number of latent variables to characterize runoff generation processes. These latent variables encapsulate the intrinsic properties of a catchment and can be inferred from catchment climate forcing and discharge data. By sampling from the latent variable space, the model generates runoff time series that closely resemble real-world observations. In this study, we trained the generative models using neural networks on data from over 3,000 global catchments and achieved prediction accuracies comparable to current deep learning models and various conventional lumped models, both within the catchments from the training set and from other regions worldwide. This suggests that the runoff generation process of catchments can be effectively captured by a low-dimensional latent representation. Yet, challenges such as equifinality and optimal determination of latent variables remain. Future research should focus on refining parameter estimation methods and exploring the physical meaning of these latent dimensions to improve model applicability and robustness. This generative approach offers a promising alternative for hydrological modeling that requires minimal assumptions about the physical processes of the catchment.

\end{abstract}

\section*{Plain Language Summary}
In this study, we use generative modeling to predict daily runoff time series from catchments in response to catchment-averaged climate forcing. Similar to how generative models synthesize human speech using features such as age and mood, our approach assumes that a few key variables can sufficiently capture the differences in runoff generation processes among catchments. These variables are not predefined by physical knowledge, but are latent variables that capture the intrinsic relationship between climate forcing and runoff time series of a catchment. The optimal values of these latent variables for a catchment can be inferred from climate forcing and runoff data. This modeling approach allows the generative model to produce realistic discharge time series in response to new climate inputs, achieving accuracy comparable to conventional process-based hydrological models. Our method provides hydrologists with a flexible tool for highly accurate streamflow prediction with minimal assumptions. Future research could focus on refining parameter estimation and exploring the physical significance of these key variables to improve model applicability and robustness.

%
%
%

\section{Introduction}

\subsection{A Generative View on Lumped Hydrological Modeling}

Rainfall-runoff modeling aims to build numerical models that can accurately predict a catchment's discharge in response to climate forcing. Among the various types of rainfall-runoff models, lumped hydrological models have a long history of development and are widely used in hydrology \cite{Beven2011}. In general, in a lumped hydrological model, the average responses of a catchment to catchment-averaged climate forcings are modeled using empirical or physically based equations \cite{Anderson2015, Yu2015, Coron2017}. The commonly used lumped models include GR4J \cite{Perrin2003}, TOPMODEL \cite{Beven2021}, and the Xinanjiang model \cite{Zhao1992}, each containing a set of predefined equations.

A lumped model typically employs several parameters to define its characteristics, such as the soil layer's storage capacity and the infiltration rate from the surface to the soil. Once the values of these parameters are set, the model can be applied to model a catchment, where it defines a mapping from climatic inputs to runoff outputs. Typically, the number of parameters is small, ranging from 1 to 20, as indicated by the models reviewed in \citeA{Knoben2019}. This parametric approach implicitly assumes that a limited number of variables can describe the diverse behavioral patterns of different catchments under climate forcing reasonably well, and that by varying the parameter values, one can produce functions suitable for runoff prediction across different catchments \cite{Beven2015}.

The approach of creating discharge prediction functions from parameter values can be seen as a form of \textit{generative modeling}, as it provides an efficient way to generate synthetic and realistic discharge data. Generative modeling is a research field in statistics and machine learning that focuses on synthesizing realistic data that resemble observed samples of a target variable \cite{Ruthotto2021}. These samples can be, for example, a set of multi-time step discharge time series from different catchments. Generative modeling can also be defined as the practice of modeling the joint probability distribution of both the target variable (e.g., discharge time series) and its associated variables (e.g., catchment physical characteristics) \cite{Kingma2019, Tomczak2022}.

\begin{figure}
    \centering
    \includegraphics[width=15 cm]{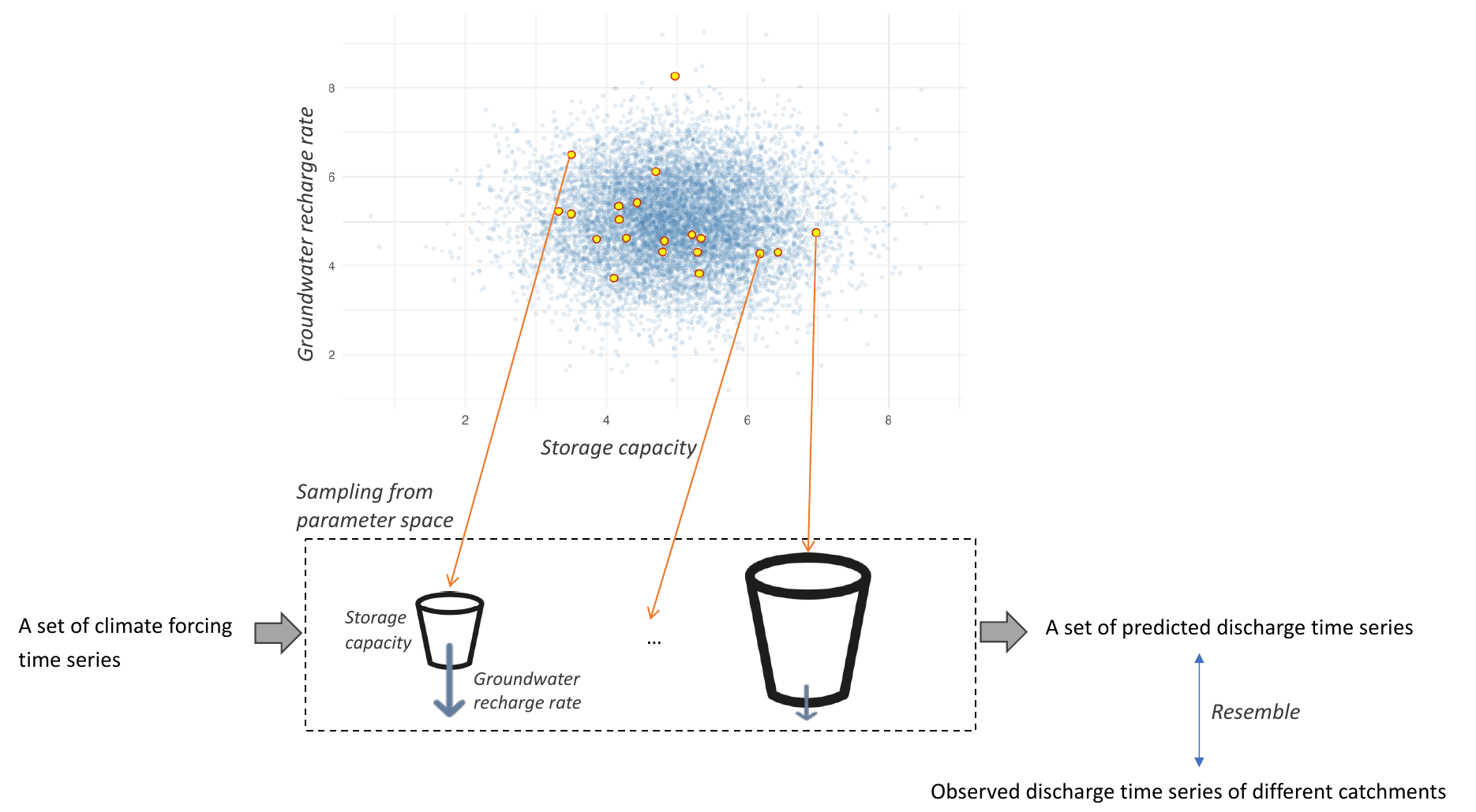}
    \caption{Illustration of the processes of generating discharge time series using a 2-parameter Leaky Bucket model.}
    \label{fig_lump}
\end{figure}

For illustration, Figure \ref{fig_lump} shows a 2-parameter Leaky Bucket model with storage capacity and constant groundwater recharge rate as its parameters. By sampling parameter values from the parameter space, different discharge prediction functions can be created. These functions can then be used to generate discharge time series that resemble those of real-world catchments.

This generative modeling approach uses hydrological theories and principles to model runoff generation processes, making it valuable for both understanding catchment hydrological processes and making accurate discharge predictions. For example, it allows the testing of different hypotheses about catchment hydrological processes through the selection of parameter values and the formulation of model equations \cite{clark2011}. In addition, it provides an efficient way to estimate the hydrological response of catchments under different climatic and environmental changes by adjusting parameter values \cite{Tian2013, Lin2014}. This approach also supports the investigation of how different parameters affect discharge patterns \cite{Song2015}, and enables model parameter regionalization for discharge prediction in ungauged catchments \cite{Gong2021}.

\subsection{Deep Learning Approaches for Generative Modeling in Hydrology}

Recently, the use of deep learning methods in rainfall-runoff modeling has gained significant traction \cite{Nearing2021, Shen2021}. These methods are used to develop runoff prediction functions that use catchment-averaged climate forcings as inputs \cite{Anderson2022}, with learning achieved by optimizing neural network parameters. Typically, deep learning models are trained on observed climate forcing and discharge data from one or more catchments. This approach differs significantly from conventional lumped hydrological models, which attempt to represent the runoff generation processes of different catchments by setting different sets of parameters \cite{Beven2015}. Notably, deep learning methods are increasingly being applied to ungauged catchments (i.e. catchments without discharge records), which requires the ability to generate functions capable of modeling arbitrary catchments by varying certain parameter values, a capability that lumped models naturally possess.

A prominent example is deep learning-based regional hydrological modeling, where the goal is to learn a general rainfall-runoff model for multiple catchments. The learned model often uses catchment characteristics (such as area and slope) to modulate model predictions \cite{Kratzert2019a, Kratzert2019b, Xu2021}. This means that a regional model takes two inputs: one is the catchment characteristics and the other is the climate forcing. These regional models can be used to model ungauged catchments if the catchment characteristics are known. Ideally, by sampling from the characteristics space, one can create runoff prediction functions that can generate realistic runoff time series that resemble real-world observations. Therefore, a regional hydrological model can be used as a generative model. Building a prediction model for multiple entities while taking into account their individual characteristics is also known as entity-aware modeling \cite{Ghosh2022, Ghosh2023}, and a regional hydrological model can be considered an entity-aware model.

Obviously, the characteristics selected in a regional hydrological model should sufficiently explain the variability in catchment behavior under climate forcing. However, it is unlikely that all contributing factors can be effectively identified and measured. This is due to our limited understanding of the physical processes and the challenges of measuring catchment physical properties at large spatial scales and in the subsurface, such as the maximum soil moisture storage capacity of the soil layer in a catchment \cite{Beven2011}.

One solution to this problem is to use \textit{latent variables} instead of measurable catchment characteristics to describe variations in catchment behavioral patterns. Latent variables are variables that cannot be directly observed. They usually have no clear physical meaning. The latent variables can be thought of as the underlying intrinsic characteristics of a catchment, and they determine or explain the differences in behavioral patterns between different catchments. The latent variables may also be referred to as implicit variables or latent factors in various literature \cite{Kingma2019, Makhzani2015, Yang2023}.

It is assumed that a general hydrological model can be built using latent variables, so that by sampling from the latent variable space one can generate runoff prediction functions capable of producing realistic runoff time series for a variety of catchments. In addition, for a given catchment, it is expected that there is at least one set of latent variable values that corresponds to a good discharge prediction function suitable for that catchment. These two assumptions are also implicitly made by conventional lumped hydrological models, where a small number of parameters explain the differences between catchments, and the existence of optimal parameter values is often assumed when optimization techniques are used to search for the optimal values.

Using latent variables to replace catchment characteristics in regional modeling can be useful as it addresses the inherent limitations in measuring and identifying all relevant factors that influence catchment behavior. This approach allows the model to account for the complex, unmeasured interactions and processes within catchments, thereby potentially improving runoff prediction accuracies. Additionally, thanks to the generative nature of this approach (which is also possessed by conventional lumped models), it enables the use of techniques developed for lumped models for various applications to better understand catchment hydrology, such as generating realistic discharge time series, parameter regionalization, and parameter sensitivity analysis. However, the usefulness of latent variable models in hydrology has not yet been fully evaluated.

Therefore, this study aims to answer the following questions:
\begin{itemize}
    \item Is it possible to use a limited number of latent variables to effectively capture the hydrological behavioral patterns of different catchments in response to climate forcing, thereby enabling the construction of general hydrological models that use these variables to provide accurate discharge predictions?
    \item Can we reliably determine the optimal latent variables for any given catchment using its climate forcing and discharge data, so that the latent variable-based model can be generalized for use with any catchment?
\end{itemize}

\section{Methods}

\subsection{Assumptions of Generative Hydrological Models} \label{sec_set_up}

\begin{figure}
    \centering
    \includegraphics[width=10 cm]{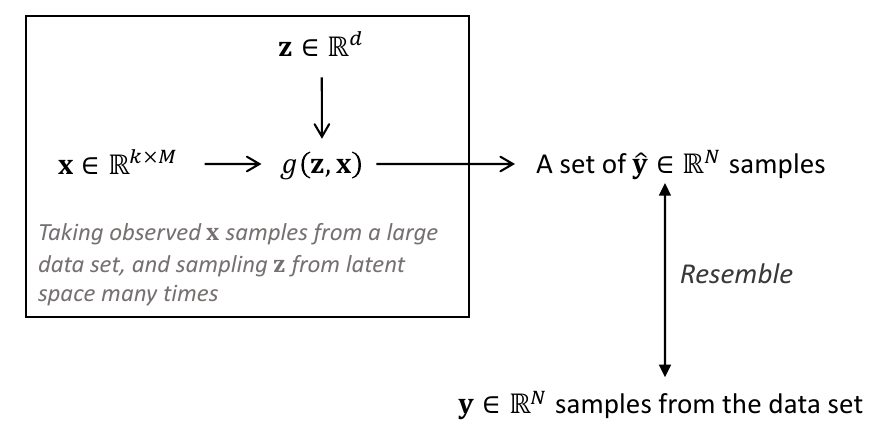}
    \caption{Illustration of the process of generating discharge time series using a generative hydrological model.}
    \label{fig_generative_framework}
\end{figure}

The goal of this research is to derive a general hydrological model \( g \) that maps a \( d \)-dimensional latent variable \( \mathbf{z} \in \mathbb{R}^d \), which describes a catchment's intrinsic hydrological characteristics, and a climate forcing time series \( \mathbf{x} \in \mathbb{R}^{k \times M} \) to a predicted catchment discharge time series \( \hat{\mathbf{y}} \in \mathbb{R}^N \). Here, \( k \) is the number of climate forcing variables considered at each time step, and \( M \) and \( N \) are the number of time steps of the forcing and discharge time series, respectively. The objective is for \( \hat{\mathbf{y}} \) to closely resemble the observed discharge time series \( \mathbf{y} \in \mathbb{R}^N \) of different catchments when sampling \( \mathbf{z} \) from some distribution \( p(\mathbf{z}) \) and feeding in appropriate \( \mathbf{x} \). Figure \ref{fig_generative_framework} illustrates the process of \( g \) generating \(\hat{\mathbf{y}}\) samples. The model \( g \) is \textit{generative} because it can effectively generate \( \hat{\mathbf{y}} \) samples that closely resemble real-world observations.

The basic assumption for defining the generative model above is that there exists a \( d \)-dimensional latent variable \( \mathbf{z} \) that can sufficiently characterize the hydrological behavioral pattern of a catchment in response to climate forcing. It is also assumed that there exists a generative model \( g \) capable of reproducing the observed discharge time series of different catchments given appropriate \( \mathbf{z} \) and \( \mathbf{x} \) values. Furthermore, it is implicitly assumed that for a given catchment \( c_i \), we can identify at least one \( \mathbf{z}_i \) from \(\mathbb{R}^d\) (i.e., \(d\) numerical values) that can be used to define a hydrological model \( g\left( \mathbf{z}_{i}, \mathbf{x} \right) \) capable of reproducing the observed discharge time series of \( c_i \). These assumptions can be verified if we can find an appropriate generative hydrological model \( g \) and the \( \mathbf{z} \) values for different catchments.

\subsection{Learning a Generative Hydrological Model}\label{sec_learning_generative}

This research employs a data-driven approach to learn a generative hydrological model \( g \) from pairs of observed climate forcing and discharge time series \( \left( \mathbf{x}, \mathbf{y} \right) \) of multiple catchments. In this study, the model \( g \) is represented by a neural network, and is denoted as \( g_\theta \), where \( \theta \) represents the parameters of the neural network.

Assume that the dataset available for the learning neural network parameter \( \theta \) is \( D = \bigcup_{i=1}^{n} \{ (\mathbf{x}_{ij}, \mathbf{y}_{ij}) \mid j = 1, 2, \ldots, n_i \} \). Here, the union symbol \( \bigcup \) is used to denote the combination of datasets from all \( n \) catchments into a single dataset \( D \). In this context, \( \mathbf{x}_{ij} \) and \( \mathbf{y}_{ij} \) are the \( j \)-th climate forcing and discharge observations for the \( i \)-th catchment, and \( n_i \) is the number of observations for the \( i \)-th catchment.

The optimal \( \mathbf{z}_i \) value associated with a catchment \( c_i \) from the dataset \( D \) can be jointly learned with \(\theta\) by solving the following optimization problem:
\begin{equation}
\min\limits_{\theta, \{\mathbf{z}_i\}_{i=1}^{n}} \sum_{i=1}^{n} \sum_{j=1}^{n_i} L\left( g_\theta\left( \mathbf{z}_{i}, \mathbf{x}_{ij} \right), \mathbf{y}_{ij} \right),
\label{eq_objective_single}
\end{equation}
where \( L \) is a loss function that measures the similarity between the observed and predicted discharge time series. Common choices for \( L \) include the mean square error (MSE) and mean absolute error (MAE). This optimization problem essentially aims to minimize the difference between the predicted and observed discharge time series by optimizing both \(\theta\) and the \( \mathbf{z}_i \) values.

Once the neural network structure (e.g., number of layers, type of each layer, number of neurons) and the model training hyperparameters (such as the learning rate and the number of epochs) have been set, common methods for training neural networks, such as backpropagation and stochastic gradient descent, can be used to solve this problem \cite{Goodfellow2016}. In particular, $d$, the dimension of the latent variable $\mathbf{z}$, is also a hyperparameter. The methods for determining the optimal network structure and other hyperparameters are described in Section \ref{sec_training_strategy}.

\subsection{Application of a Generative Model to Arbitrary Catchments}\label{sec_application_to_arbitrary}

As mentioned in Section \ref{sec_set_up}, we assume that for a given generative hydrological model \( g \) and an arbitrary catchment \( c_i \), there exists a set of latent variable values \(\mathbf{z}_{i}\) that define accurate discharge prediction functions \( g\left( \mathbf{z}_{i}, \mathbf{x} \right) \) for \( c_i \). Note that \( c_i \) can be an arbitrary catchment that is not necessarily from the dataset \( D \) used to train \( g \). 

These optimal \(\mathbf{z}_{i}\) values can generally be derived in a manner similar to conventional lumped models (which can also be viewed as generative models). The first method involves using a catchment's physical properties to infer the optimal parameter values, a process known as ungauged modeling. This method is useful when catchment discharge data is not available. The second method involves searching through the latent space using optimization algorithms to minimize the discharge prediction error of the resulting model \( g\left( \mathbf{z}_{i}, \mathbf{x} \right) \) for \( c_i \), which requires using the actual discharge records. This process is known as model calibration or parameter optimization.

The effectiveness of the model calibration method for generative models is evaluated in this study. In such a method, the parameter estimation problem can be formulated as an optimization problem. For a given catchment \( c_i \) with \( n_i \) samples of climate forcing and discharge pairs, the optimal \(\mathbf{z}_{i}\) value can be found by solving the following optimization problem:
\begin{equation}
    \min\limits_{\mathbf{z}_{i}} \sum_{j=1}^{n_i} L\left( g_\theta\left( \mathbf{z}_{i}, \mathbf{x}_{ij} \right), \mathbf{y}_{ij} \right),
\label{eq_optimize_z_i}
\end{equation}
where \( L \) is a loss function that measures the difference between the observed and predicted discharge time series, and \(\left( \mathbf{x}_{ij}, \mathbf{y}_{ij}\right)\) is a pair of observed climate forcing and discharge time series, with \( j \) being the index of the observations. In general, the commonly used model calibration methods in hydrology are applicable for solving these optimization problems, such as the Shuffled Complex Evolution (SCE-UA) algorithm \cite{Duan1992} and genetic algorithms (GA) \cite{Katoch2021}. 

The quality of the derived optimal parameters can then be further verified using an additional set of climate forcing and discharge observations, called a validation set or test set.

\section{Experimental Setup}

This study designs numerical experiments to validate whether the assumption of latent variable representation of catchment characteristics (as discussed in Section \ref{sec_set_up}) is useful for building generative hydrological models capable of providing accurate discharge predictions. The ability of the resulting model to provide accurate predictions will be tested in two ways.

\begin{enumerate}
    \item This study first uses a global catchment climate and discharge dataset to train a generative model \( g_\theta \), and then evaluates \( g_\theta \)'s ability to provide accurate discharge predictions for catchments within the dataset using climate and discharge records from another time period. Thus, this experiment is conducted in an in-sample testing or regional modeling setting.
    \item This study then evaluates the usefulness of \( g_\theta \) in modeling catchments outside the training dataset, where the optimal model parameters are obtained through a model calibration procedure (as described in Section \ref{sec_application_to_arbitrary}). Thus, this experiment is an out-of-sample testing or model calibration study. \( g_\theta \)'s performance in model calibration studies is also compared to conventional process-based lumped hydrological models.
\end{enumerate}

The setup of numerical experiments is described next.

\subsection{Neural Network Architecture of the Generative Model}

In this study, a generative model is implemented as a neural network \(g_\theta\), where \(\theta\) denotes the parameters of the network. The model \(g_\theta\) accepts two inputs: the climate forcing time series \(\mathbf{x} \in \mathbb{R}^{3 \times 730}\) and a latent variable \(\mathbf{z} \in \mathbb{R}^d\). (1) \(\mathbf{x}\) is a multi-dimensional dataset consisting of two years of daily observations for precipitation (P), temperature (T), and potential evapotranspiration (PET). (2) \(\mathbf{z}\) is a \(d\)-dimensional variable that captures underlying catchment hydrological characteristics. The optimal value of \(d\) is identified through a hyperparameter optimization process, as detailed in Section \ref{sec_training_strategy}.

The output of \(g_\theta\) is a time series \(\hat{\mathbf{y}} \in \mathbb{R}^{365}\), representing the predicted daily runoff depth over the catchment area for the second year covered by \(\mathbf{x}\). Essentially, \(g_\theta\) uses the two-year climate data to predict the discharge for the second year.

\begin{figure}
\centering
\includegraphics[width=14.5 cm]{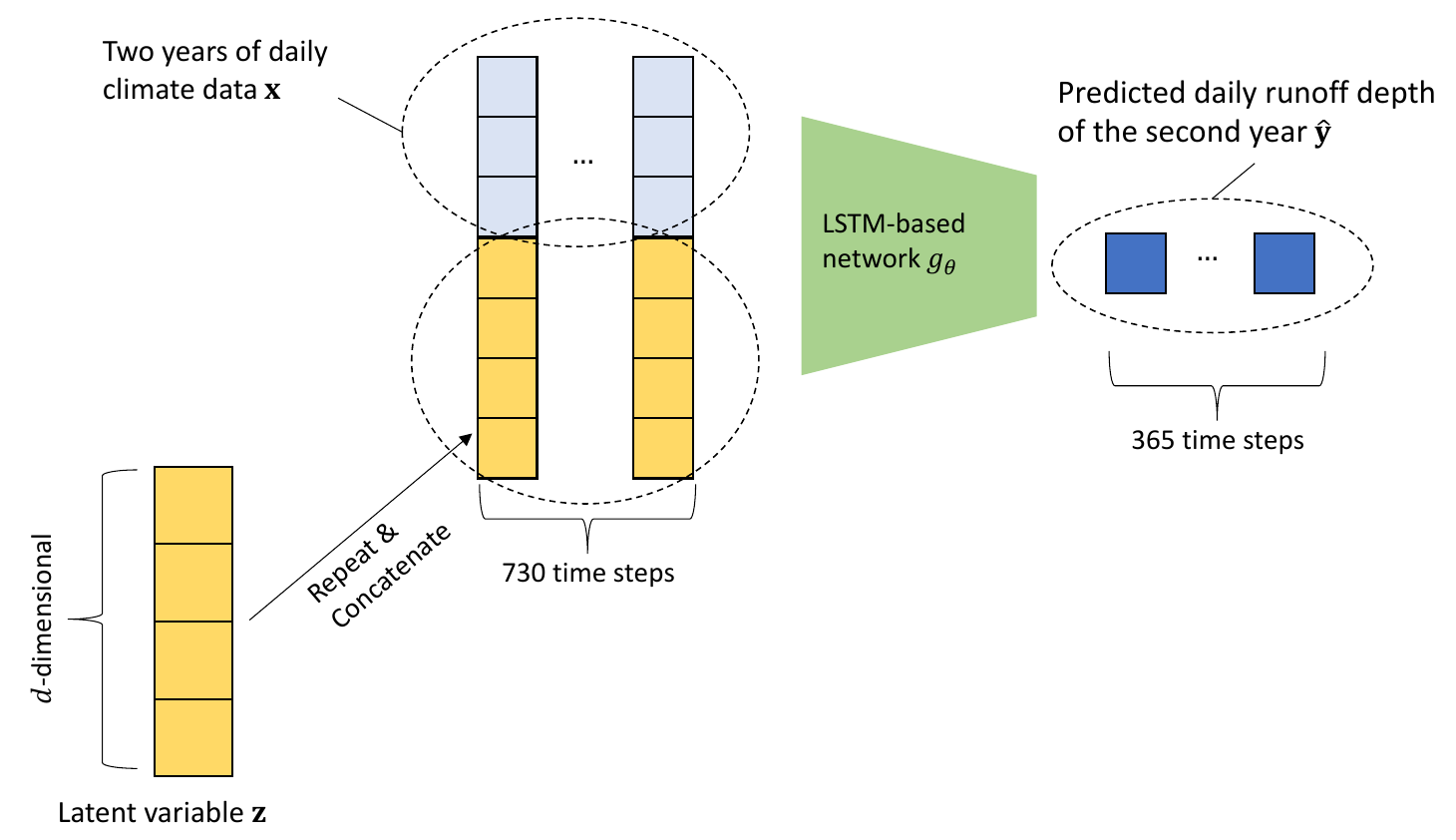}
\caption{Architecture of the generative hydrological model \(g_\theta\) used in this study.}
\label{fig_architecture}
\end{figure}

The network architecture of \(g_\theta\) is shown in Figure \ref{fig_architecture}, which can be considered as an auto-decoder model \cite{Park2019}. At each time step, the latent variable \(\mathbf{z}\) is concatenated with the climate forcing data for that time step, and this combined input is fed into a Long Short-Term Memory (LSTM) network \cite{Hochreiter1997,Gers2000}. The LSTM processes the input sequentially, producing an output at each time step. This output is then further refined by a multilayer perceptron (MLP), i.e, a series of densely connected layers. The final 365 values predicted by the network form the discharge time series \(\hat{\mathbf{y}}\). The optimal number of layers and neurons in the network are determined through a hyperparameter optimization process, as discussed in Section \ref{sec_training_strategy}.

For an introduction to LSTM in rainfall-runoff modeling, see \citeA{Kratzert2018}. It is also possible to use other neural network architectures for time series modeling in \(g_\theta\), such as EA-LSTM \cite{Kratzert2019a}, Transformers \cite{Vaswani2017,Liu2024}, and Temporal Convolutional Networks (TCN; \citeA{Bai2018}).

\subsection{Training and In-Sample Evaluation Strategy}\label{sec_training_strategy}

Training the final generative hydrological model \(g_\theta\), as described in Section \ref{sec_learning_generative}, involves setting several hyperparameters. These include, e.g., the dimension of the latent variables \(d\), the number of layers in the LSTM network, the learning rate, batch size, and the number of epochs. To determine the optimal hyperparameters, this study employs an optimization approach that involves training models on a training dataset and validating them on a validation set. The objective is to experiment with different hyperparameter values to minimize prediction errors as measured by mean squared error (MSE).

In the hyperparameter optimization process, the climate forcing-discharge time series dataset of multiple catchments \(D\) is divided into training, validation, and test sets, with specific periods allocated to each set. Each catchment in \(D\) is assigned a fixed latent variable value \(\mathbf{z}_i\), which is learned alongside the network parameters \(\theta\). The \(\mathbf{z}_i\) values are initialized randomly before training, and no restrictions are placed on their distributions. This means that we impose no constraints on the distribution of latent variables \(p\left( \mathbf{z} \right)\), which is a simpler assumption compared to some popular latent variable-based generative models, such as Variational Autoencoders \cite{Kingma2019} and DeepSDF \cite{Park2019}. These models typically constrain \(p\left( \mathbf{z} \right)\) to simple distribution functions, such as normal distributions, to facilitate easy sampling from latent space. As will be shown in the results section, this simpler assumption is sufficient for hydrological modeling in the context of this study.

This study utilizes a Bayesian optimization algorithm implemented in the ``Optuna" Python library \cite{Akiba2019} to solve the hyperparameter optimization problem, due to its high efficiency in finding solutions with fewer iterations. The candidate values considered for \(d\) are 2, 4, 8, and 16, reflecting the typically low number of parameters used in conventional lumped hydrological models, as shown by the models examined in \citeA{Knoben2020}. This assumption is based on the idea that a small number of variables is sufficient to characterize a catchment's hydrological characteristics in conventional lumped models. The ranges for the other hyperparameters and the setting of the Bayesian optimization algorithm are provided in Supporting Information Table S1.

Once the optimal hyperparameters are identified, the model is trained on the combined training and validation sets and tested on the test set to estimate the generalization error in terms of the Kling-Gupta Efficiency (KGE) criterion scores \cite{Gupta2009}. Finally, the final \(g_\theta\) model is trained on the entire dataset \(D\).

\subsection{Out-of-Sample Evaluation Strategy}\label{sec_out_of_sample_evaluation_strategy}

After training the generative model \(g_\theta\), its ability to model out-of-sample catchments—those not included in the training dataset \(D\)—is evaluated. As outlined in Section \ref{sec_application_to_arbitrary}, for an arbitrary catchment \(c_i\) with available climate forcing and discharge records, a generic optimization algorithm can be utilized to derive the optimal latent variable values \(\mathbf{z}_i\) for \(g_\theta\). In this study, we employ a genetic algorithm (GA) implemented in the ``PyGAD" Python library \cite{Gad2021} to determine these optimal values that maximize the KGE scores. Details on the GA settings are provided in the supporting document S1, and an introduction to GA can be found in \citeA{Luke2013}.

For each catchment \(c_i\), the climate forcing and discharge data are divided into a calibration period and a testing period. The GA is used to derive the optimal \(\mathbf{z}_{i}\) value that minimizes prediction error during the calibration period. This optimal \(\mathbf{z}_{i}\) is then input into \(g_\theta\) to evaluate the prediction error during the test period.

\subsection{Data}

\subsubsection{Catchment Climate Forcing and Discharge Data}

This study utilizes three catchment hydrology datasets from various regions worldwide: CAMELS (Catchment Attributes and Meteorology for Large-sample Studies; \citeA{Newman2014_data,Newman2015,Addor2017}), CAMELS-DE Version 0.1.0 \cite{Dolich2024,Loritz2024}, and Caravan Version 1.4 \cite{Kratzert2023}. CAMELS and CAMELS-DE cover catchments from the US and Germany, respectively, while Caravan includes catchments globally. Catchments were selected from each dataset based on their location and data availability. To evaluate the model \(g_\theta\)'s ability to generalize across regions, only non-US catchments were selected from the Caravan dataset for training. Subsequently, the model was tested on US and German catchments from the CAMELS and CAMELS-DE datasets. Notably, the versions of Caravan and CAMELS-DE used in this study contain minimal overlap in terms of shared catchments.

Catchment mean daily precipitation (P), temperature (T), and potential evapotranspiration (PET) were used as climate forcing variables due to their common application in many conventional lumped models \cite{Knoben2019}. Daily runoff (Q) served as the prediction target. These four variables were available in the Caravan and CAMELS-DE datasets and were derived from the original CAMELS dataset by \citeA{Knoben2020}. For inclusion in the study, each catchment was required to have at least two years (not necessarily continuous) of complete climate forcing and discharge records.

\begin{figure}
\centering
\includegraphics[width=14.5 cm]{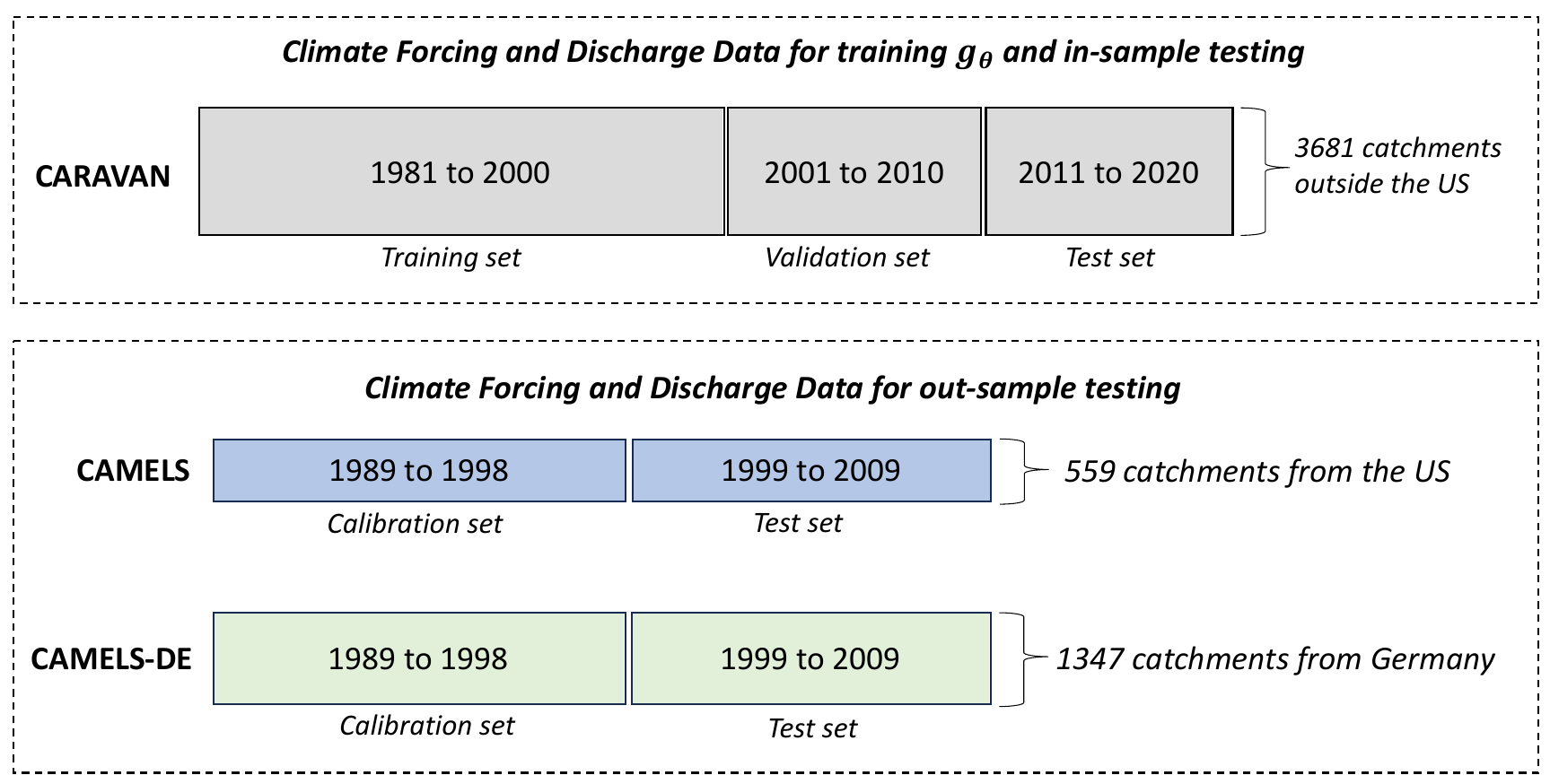}
\caption{Roles of datasets and dataset splitting strategy.}
\label{fig_data_split}
\end{figure}

The roles of the datasets and dataset division for training, calibration, validation, and testing of \(g_\theta\) are illustrated in Figure \ref{fig_data_split}. For the Caravan dataset, three subsets were created to train and test the generative models using the methods described in Sections \ref{sec_learning_generative} and \ref{sec_training_strategy}. Both the CAMELS and CAMELS-DE datasets were divided into two subsets: a calibration set and a test set. The optimal \(\mathbf{z}_i\) values of different catchments were derived and tested using the calibration and test sets, as outlined in Sections \ref{sec_application_to_arbitrary} and \ref{sec_out_of_sample_evaluation_strategy}.

\subsubsection{Benchmark Data on Lumped Hydrological Models' Performance}

\citeA{Knoben2020} calibrated 36 types of lumped hydrological models for 559 catchments from the CAMELS dataset. These models include the Xinanjiang Model, TOPMODEL, HBV-96, VIC, and GR4J. They reported the discharge prediction accuracies of the calibrated models, including the KGE scores \cite{Gupta2009}. The model performance data is available with their publication, providing a comprehensive benchmark for evaluating hydrological models. More details about the model performance dataset can be found in Knoben et al. (2020). In this study, these KGE scores are used as benchmarks to assess the effectiveness of the generative models in discharge prediction tasks.

\section{Numerical Experiments and Results}

This study begins by evaluating the predictive performance of generative hydrological models \(g_\theta\) in both in-sample and out-of-sample settings. It then briefly examines the distribution of optimal \(\mathbf{z}_i\) values for a given catchment and how this value affects discharge predictions.

\subsection{In-Sample Testing}\label{sec_in_sample_results}

\subsubsection{Experiments Performed}

In the first experiment, a generative hydrological model \(g_\theta\) was trained using the combined training and validation sets from the Caravan dataset. During this training process, the optimal latent variable values \(\mathbf{z}_i\) for the 3,681 catchments were also determined. See Sections \ref{sec_learning_generative} and \ref{sec_training_strategy} for details. The resulting discharge prediction models, represented as \(g_\theta\left( \mathbf{z}_{i}, \mathbf{x}_{ij} \right)\), were then evaluated on the test set of Caravan, which constitutes an in-sample testing or regional modeling scenario.

\subsubsection{Results}

\begin{figure}
\centering
\includegraphics[width=10 cm]{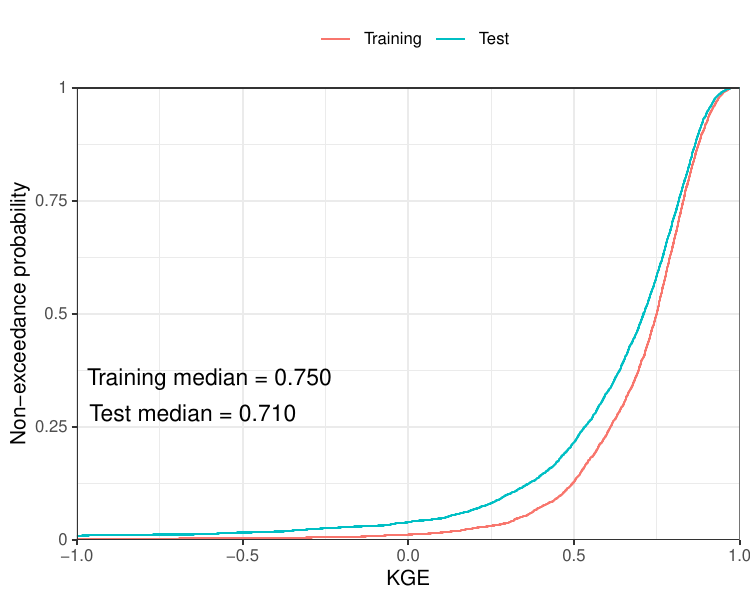}
\caption{Non-exceedance probability of the training and test KGE scores for each Caravan catchment. The medians of the training and test KGE scores are also displayed. For clarity, results for KGE values less than -1 are not shown here but can be found in Figure \ref{fig_caravan_train_vs_test}. The test KGE scores for two catchments unavailable because the recorded runoffs were constantly 0.}
\label{fig_ecdf}
\end{figure}

Through hyperparameter optimization, the optimal dimension \(d\) for the latent variables was found to be 8. Consequently, for each catchment, an 8-dimensional latent variable \(\mathbf{z}_i\) was derived to represent its hydrological characteristics. The non-exceedance probability of the KGE scores for the model during the training and test periods is illustrated in Figure \ref{fig_ecdf}. Note that the training period here includes the timeframes specified by both the training and validation sets of the Caravan dataset, as both were used for training \(g_\theta\).

The median KGE scores for the training and test periods are 0.750 and 0.710, respectively. This indicates that the prediction accuracy was generally satisfactory, although a performance gap between the training and test periods is evident.

\begin{figure}
\centering
\includegraphics[width=16 cm]{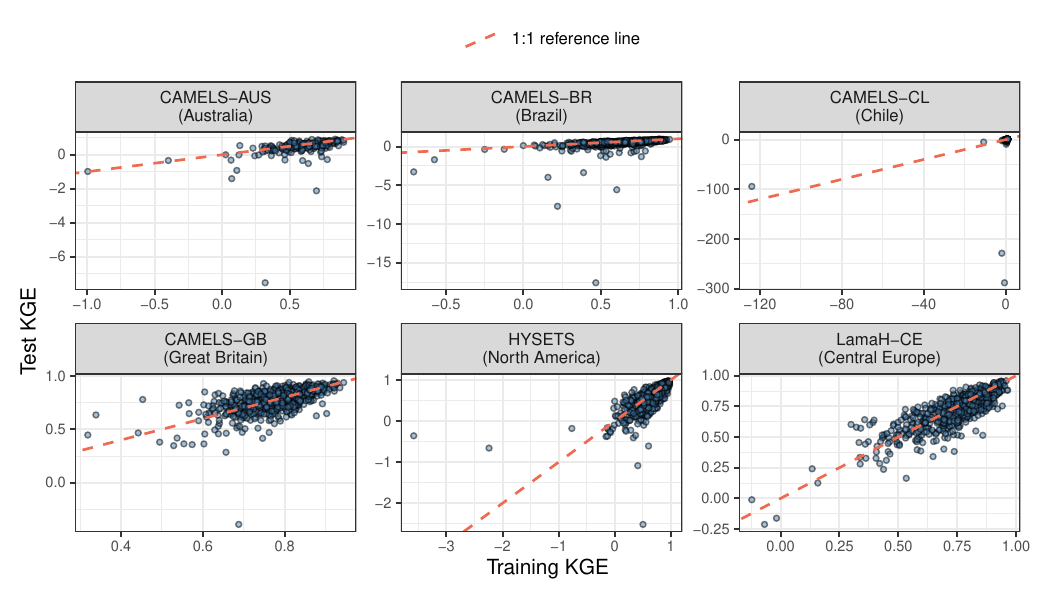}
\caption{Comparison of training and test KGE scores for catchments from different countries or regions covered by the Caravan dataset. The label of each subplot indicates the name of the original dataset from which the Caravan dataset extracts the data. The subplot label also indicates the main country or region covered by each original dataset. The test KGE scores for two catchments unavailable because the recorded runoffs were constantly 0.}
\label{fig_caravan_train_vs_test}
\end{figure}

We further examined the generative model's performance across catchments from different regions. Figure \ref{fig_caravan_train_vs_test} provides a comparison of the training and test KGE scores for model instances across different countries and regions covered by the Caravan dataset. The figure shows a positive correlation between the training and test KGE scores for the regions considered. For catchments in Great Britain and Central Europe, both training and test KGE scores are generally higher than 0, indicating robust model performance. In contrast, some regions exhibit poor performance, as evidenced by a few large negative KGE scores.

\subsubsection{Discussions}

The test KGE results shown in Figures \ref{fig_ecdf} and \ref{fig_caravan_train_vs_test} indicate that the generative model trained solely on climate forcing and runoff data was generally effective in an in-sample test or regional modeling setting. Unlike most current regional modeling studies, this approach did not utilize catchment attributes that describe the physical characteristics of a catchment. The proposed generative modeling framework does not assume a specific correlation between catchment attributes and hydrological functions. Instead, it focuses on learning a generative model and a small number of latent variables capable of accurately reproducing observed hydrographs.

The optimal number of latent dimensions \(d\) was found to be 8, suggesting that the hydrological behavioral characteristics of different catchments can be effectively captured using a small number of numerical values. This finding aligns with the results of \citeA{Yang2023}, which indicate that the degree of fit between a catchment and a hydrological model (whose parameter values are set) can be sufficiently explained by a limited number of factors. However, the specific information encoded in each dimension is unknown. This contrasts with conventional lumped model classes, where the physical meaning of parameters is typically  understood.

Similar to conventional process-based lumped models, the generative model is also susceptible to overfitting, as evidenced by the performance gap between the training and test KGE scores shown in Figures \ref{fig_ecdf}. This overfitting may result from catchments exhibiting different runoff generation patterns during the training and test periods, leading model instances to learn ``noise" in the forcing and runoff time series alongside the true hydrological patterns. 

The overfitting problem may be caused by learning noise from noisy data, which could potentially be mitigated by implementing regularization methods and incorporating more extensive training data. Since this study aims to provide a straightforward introduction to the generative modeling approach, these options were not explored here. Furthermore, the factors contributing to the differences in predictive performance across catchments were not investigated in this work.

\subsection{Out-of-Sample Testing}\label{sec_out_of_sample_results}

\subsubsection{Experiments Performed}

We conducted a second set of experiments to evaluate the generative model \(g_\theta\) in an out-of-sample context. In these experiments, the model was trained on all available data from the selected Caravan catchments using the optimal hyperparameter values identified in Section \ref{sec_in_sample_results}. The trained model \(g_\theta\) was then tested on selected CAMELS and CAMELS-DE catchments, representing an out-of-sample testing scenario. Here, the optimal latent variable \(\mathbf{z}_i\) values of different catchments were derived using the calibration sets and validated on the test sets. 

For the CAMELS catchments, the data used for calibration and testing were the same as those used for evaluating the lumped models in \citeA{Knoben2020}. This setup allowed for a direct comparison of the predictive accuracy between the generative model and the lumped models calibrated in \citeA{Knoben2020}.

\subsubsection{Results}

\begin{figure}
    \centering
    \includegraphics[width=16 cm]{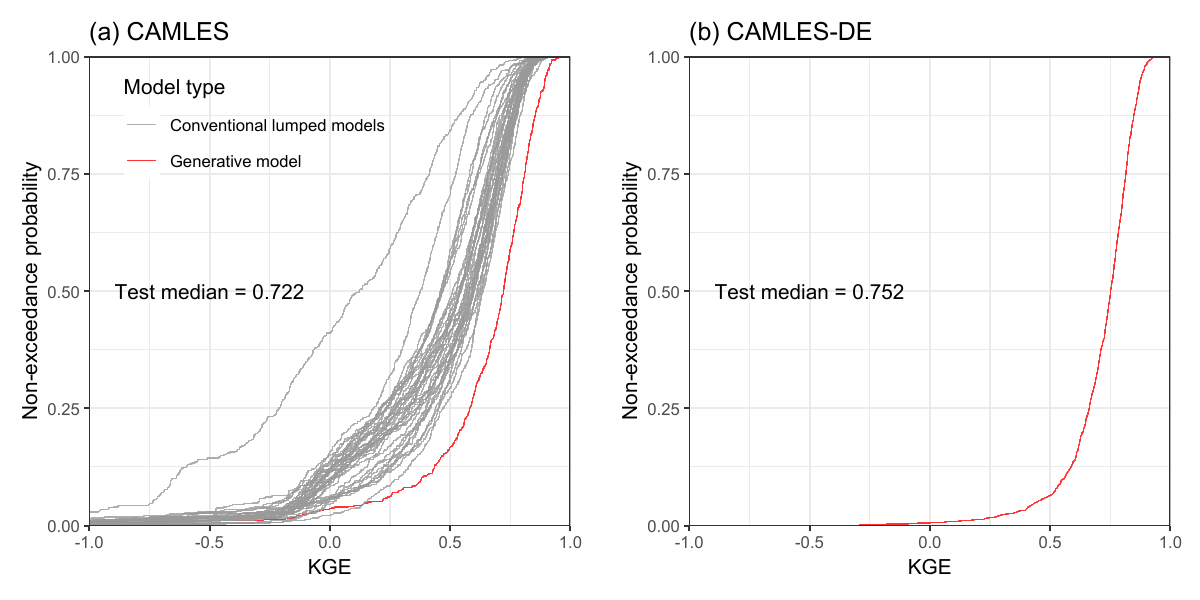}
    \caption{The non-exceedance probability of the test KGE score for the 36 types of conventional lumped models and the generative model obtained for (a) CAMELS catchments and (b) CAMELS-DE catchments. For clarity, results for KGE values less than -1 are not shown in (a). The lumped model instances were calibrated in \citeA{Knoben2020}.}
    \label{fig_generative_vs_lumped}
\end{figure}

The predictive performance of the generative model in out-of-sample testing scenarios is shown in Figure \ref{fig_generative_vs_lumped}. The median test KGE scores were 0.722 and 0.752 for the CAMELS and CAMELS-DE catchments, respectively. Notably, the median KGE score for the CAMELS catchments was comparable to previous deep learning studies conducted in a regional modeling setting (i.e., training and evaluating the model on the same set of catchments), such as those by \citeA{Botterill2023} and \citeA{Li2022}. 

Figure \ref{fig_generative_vs_lumped}a shows a comparison between the predictive performance of the generative model and 36 types of conventional process-based lumped models on the 559 CAMELS catchments. The generative model generally outperformed these conventional lumped models, with the exception of the bottom 10\% of the catchments.

\begin{figure}
    \centering
    \includegraphics[width=12 cm]{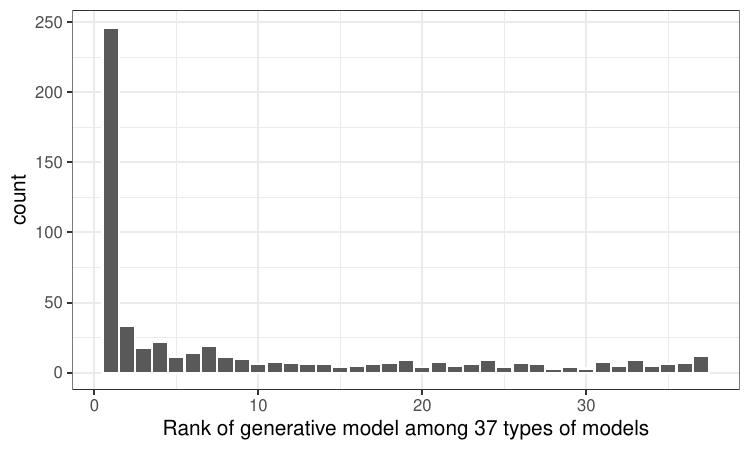}
    \caption{Histogram of the ranking of the generative model when ranked with the corresponding 36 types of conventional process-based lumped models in the 559 CAMELS catchments. The test KGE scores were used in the ranking. The lumped models were calibrated in \citeA{Knoben2020}.}
    \label{fig_rank_histogram}
\end{figure}

For each CAMELS catchment, the performance of the generative model was compared to the corresponding lumped models, and the ranking of the generative model is shown in Figure \ref{fig_rank_histogram}. The generative model ranked in the top 1, top 3, and top 10 for 44.0\%, 53.1\%, and 69.8\% of the catchments, respectively. However, it is important to note that these models can have very poor rankings in some catchments, indicating variability in performance across different regions.

\begin{figure}
    \centering
    \includegraphics[width=10 cm]{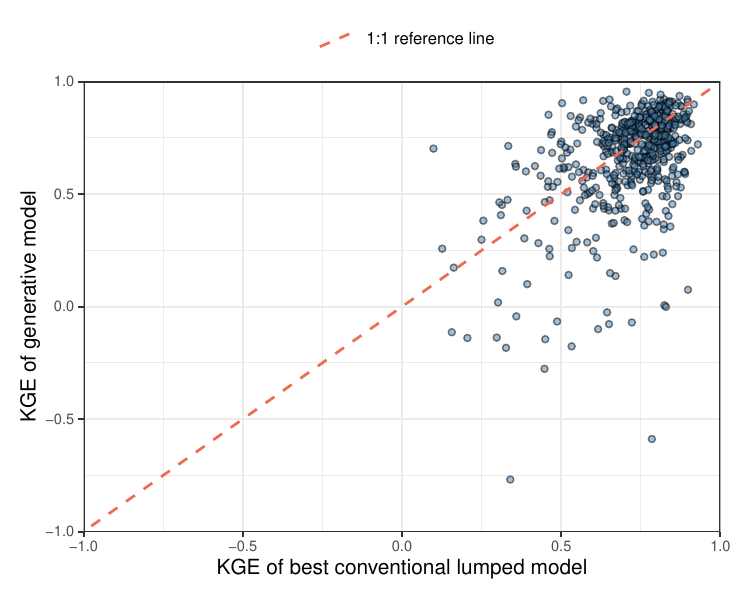}
    \caption{The test KGE scores of the generative model at the 559 CAMELS catchments compared to the best of the 36 types of lumped models. For clarity, results for KGE values less than -1 are not shown. The lumped models were calibrated in \citeA{Knoben2020}.}
    \label{fig_best_lump_vs_generative}
\end{figure}

The prediction accuracy of the generative model for each of the 559 CAMLES catchments was compared to the best of the corresponding 36 types of lumped models, as shown in Figure \ref{fig_best_lump_vs_generative}. A weak to moderate positive correlation (Pearson correlation coefficient \(r=0.450\)) was observed between the KGE score of the two sets of models. In some catchments, the generative model performed significantly worse than the best lumped models, as indicated by negative KGE scores. Notably, the KGE scores of the best lumped models were rarely negative, highlighting their relative robustness in these scenarios.

\begin{figure}
    \centering
    \includegraphics[width=14 cm]{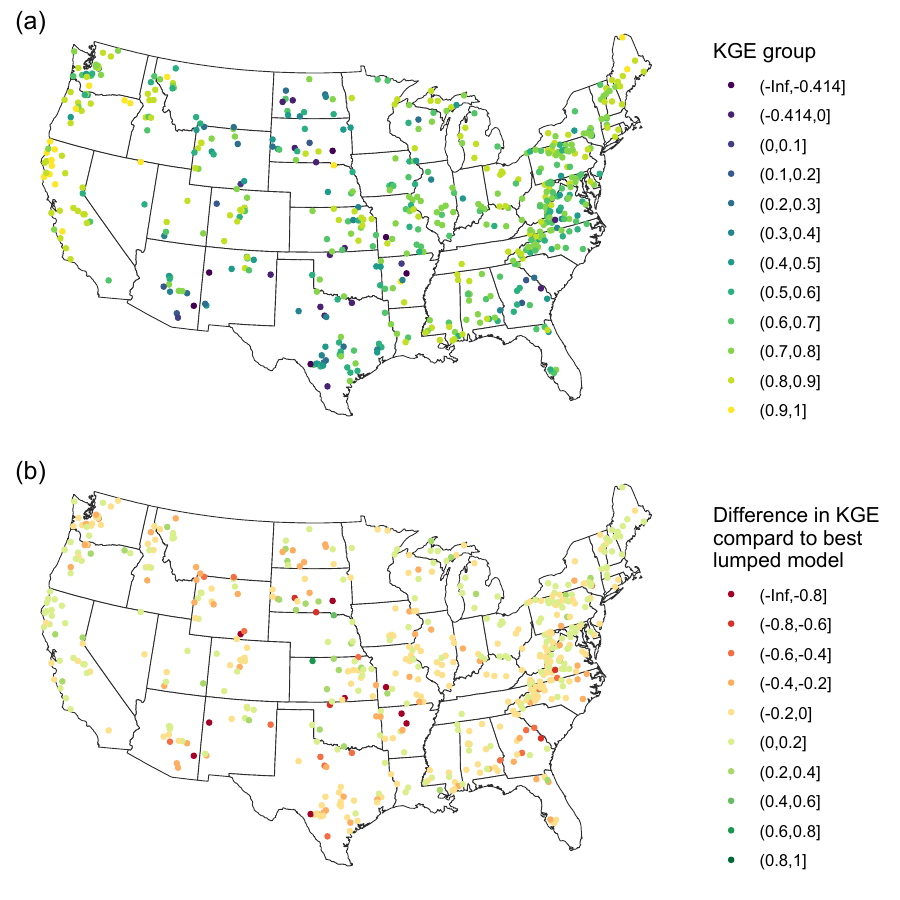}
    \caption{Spatial distribution of (a) the test KGE scores of the generative model and (b) the difference in KGE score between the generative model and the best of the 36 types of lumped models in each of the 559 CAMELS catchments. The lumped model instances were calibrated in \citeA{Knoben2020}.}
    \label{fig_KGE_map}
\end{figure}

The spatial distribution pattern of test KGE score for the generative model at the CAMELS catchments is shown in Figure \ref{fig_KGE_map}a. The model generally shows good predictive performance in the Pacific Mountain Systems and Appalachian Highlands regions, which correspond to the west and east coasts of the US. According to \citeA{Knoben2019b}, a KGE score of \(\mbox{KGE}=1-\sqrt{2}\approx-0.414\) is achieved when mean streamflow is used as a predictor. Catchments with \(\mbox{KGE}\leq-0.414\) were predominantly found in the Interior Plains, Rocky Mountain System, and Intermontane Plateaus, which are located in the central and western US.

The spatial distribution pattern of the difference between the generative model and the best lumped models, in terms of test KGE values, is shown in Figure \ref{fig_KGE_map}b. The results indicate that these two sets of model instances generally had comparable performance in the Pacific Mountain Systems and Appalachian Highlands regions. However, the generative model could perform significantly worse in other regions.

\subsubsection{Discussion}

The experimental results demonstrate that it is feasible to develop a general or generative hydrological model directly from catchment climate forcing and discharge data. By sampling from its parameter space, specifically the latent variable space, we can create discharge prediction models that produce discharge time series that resemble real-world catchments. Unlike conventional process-based lumped models, generative models are automatically learned from data without relying on specific hydrological knowledge or assumptions about catchment processes. Thus, this study introduces a data-driven approach to defining meaningful streamflow prediction models. Notably, unlike many current deep learning-based regional modeling studies, this approach does not require catchment attributes to define a model.

The generative model demonstrated predictive performance in out-of-sample testing scenarios comparable to models learned directly from data using deep learning methods (applied in in-sample testing or regional modeling scenarios), but with significantly less effort. Building a discharge prediction model for a catchment using the generative approach requires only the estimation of a few parameters (i.e., the latent variable value) using a generic calibration algorithm. In contrast, direct learning approaches require learning many more parameters, often tens of thousands. Reducing the number of trainable parameters in a model can potentially help mitigate overfitting, especially when data are limited. However, this assumption requires formal verification.

In comparison to the 36 conventional lumped models, the generative model showed comparable or better ability to model the runoff generation process in the majority of the CAMELS catchments. Therefore, the generative models can be a useful addition to a modeler's lumped modeling toolbox. The better performance of the machine learning models in some catchments indicates that our hydrological knowledge (used to develop lumped model classes) was insufficient, which is similar to the conclusions of other machine learning research in hydrological modeling. However, it should be noted that the generative model can have poor performance in some catchments where the conventional lumped models can have good performance. The poor performance may be explained by that the training data set does not contain catchments with similar runoff generation characteristics. It is also possible the ability of the generative model to extrapolate different runoff generation mechanisms is limited. Future studies can therefore use a larger or more diverse data set for training of a generative model and then test whether this model has a broader applicability.

The comparison between the best process-based lumped model and the generative model shows that a collection of conventional lumped models can deliver competitive results compared to modern data-driven approaches such as deep learning. Notably, the lower bound of their performance is often higher than that of data-driven models, underscoring the continued utility of lumped models in hydrological prediction tasks. To thoroughly assess the effectiveness of both lumped and data-driven models under various conditions, rigorous testing using methods such as metamorphic testing \cite{Yang2021} and interpretable machine learning techniques \cite{Molnar2022} is recommended.

\subsection{Estimation of Latent Variable Values}\label{sec_latent_variable_results}

\subsubsection{Experiments Performed}

This section describes an experiment conducted to investigate the challenge of determining the optimal latent variable \(\mathbf{z}_i\) values for a given catchment and to assess the impact on predicted discharges. For illustration, we selected an arbitrary catchment from the CAMELS dataset: the Fish River catchment near Fort Kent, Maine (USGS gage: 01013500). We performed parameter optimization ten times using the method described in Sections \ref{sec_application_to_arbitrary} and \ref{sec_out_of_sample_evaluation_strategy} during the calibration period to find ten sets of optimal \(\mathbf{z}_i\) values. These different versions of the \(\mathbf{z}_i\) values were then compared in terms of their values, the discharge prediction accuracy during the test period of the resulting model \(g_\theta \left(\mathbf{z}_i, \mathbf{x}\right)\) , and the shape of the predicted discharge time series.

\subsubsection{Results}

\begin{figure}
    \centering
    \includegraphics[width=12 cm]{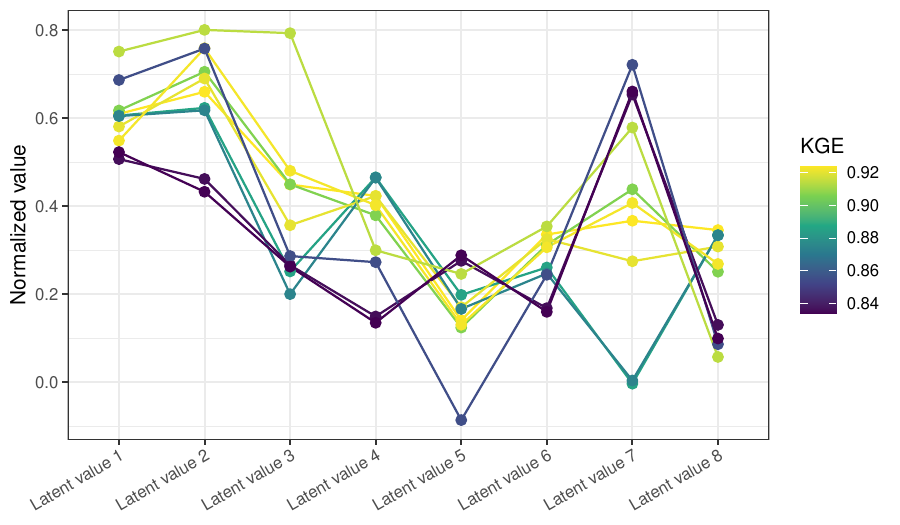}
    \caption{The normalized values of the optimal latent variable \(\mathbf{z}_i\) values for Fish River near Fort Kent, Maine (USGS gage 01013500) in the generative model, derived from ten genetic algorithm optimization runs. The corresponding test KGE scores are also shown.}
    \label{fig_latent_variable_distribution}
\end{figure}

The ten versions of the optimal \(\mathbf{z}_i\) values, derived from ten optimization runs for the Fish River catchment, are normalized and compared in Figure \ref{fig_latent_variable_distribution}. These values are normalized based on the initial upper and lower bounds for each dimension, which define the range from which the initial population was selected. The values of the upper bound are normalized to one, and the values of the lower bound are normalized to zero. It is important to note that the optimal solutions may extend beyond these bounds due to the random mutation method employed. The corresponding test KGE scores for these \(\mathbf{z}_i\) values are also shown in the figure. Generally, the \(\mathbf{z}_i\) values associated with good KGE scores are relatively close to each other, while those with poorer KGE scores exhibit larger deviations from the high-scoring values. These larger deviations are observed across all eight latent variable dimensions.

\begin{figure}
    \centering
    \includegraphics[width=16 cm]{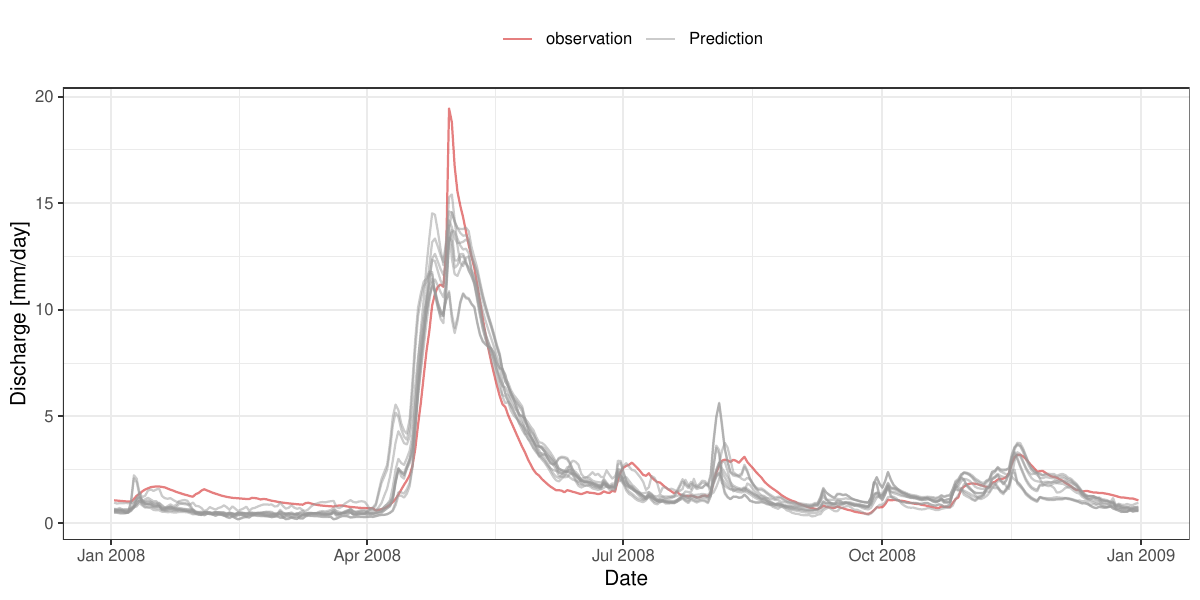}
    \caption{The predicted hydrographs of the Fish River catchment near Fort Kent, Maine (USGS gage 01013500) using the optimal latent variable \(\mathbf{z}_i\) values obtained from ten genetic algorithm optimization runs.}
    \label{fig_predicted_hydrograph}
\end{figure}

The predicted hydrographs using the ten versions of \(\mathbf{z}_i\) values are compared with the observed hydrograph in Figure \ref{fig_predicted_hydrograph}. The figure shows a one-year hydrograph with the largest runoff event during the test period. The predicted hydrographs resulting from the ten sets of \(\mathbf{z}_i\) values are clearly different. At certain times, the predicted hydrographs show similar patterns in their relationship to the observed hydrograph; for example, all predictions tend to underestimate the flood peak and overestimate the flow during the recession limb of the hydrograph.

\subsubsection{Discussions}

The experimental results on the differences between the ten sets of optimal latent variable values indicate that it can be challenging to determine a single optimal set of parameter values in a generative model. Similar to conventional process-based models, there is equifinality \cite{Beven2006}, meaning that different sets of parameter values can all provide acceptable predictions, making it difficult to identify the best one. The parameters associated with high prediction accuracy tend to cluster closely together, suggesting that effective parameter sets may reside within a small subregion of the parameter space.

The variations in the model's performance, as measured by test KGE scores, suggest that the genetic algorithm (GA) applied in this study may not reliably identify the best parameters. Future studies could explore more advanced methods to optimize the parameters, similar to approaches used in conventional lumped models. It is important to note that in this study, we used a single set of hyperparameter values for the GA for all catchments considered, and future research could involve optimizing these hyperparameters for each catchment individually.

Unlike conventional lumped model classes, the ``meaning" of each dimension of the latent variable (i.e., each parameter) is automatically learned from the data, rather than being predefined. It would be interesting to evaluate whether the optimal latent vector of a catchment can be predicted by its physical properties. If so, a reasonable model could be created for an ungauged catchment using the generative model and the catchment's physical properties. The optimal latent variable values of different catchments may also be useful for catchment classification and catchment similarity analysis research. Further investigation into these topics is recommended. Additionally, sensitivity analysis methods currently used in hydrology \cite{Song2015} may also be valuable for studying how the latent vector values affect the predicted hydrograph.

The ten versions of latent variable values correspond to different predicted hydrographs, though they show some agreement with the observed hydrograph. However, predictions for peak flow are noticeably different. Therefore, before the generative model can be applied in real-world applications, its parameter identification problem needs to be thoroughly studied to explore possible solutions and understand its impact on prediction accuracy in different contexts, such as prediction of extreme events or prediction under climate change scenarios.

\section{Conclusions}

This study presents a generative modeling approach to rainfall-runoff modeling, where models are automatically learned from catchment climate forcing and discharge data, using simple assumptions about the limited number of variables needed to characterize a catchment's runoff generation process. The learned models are capable of producing discharge time series that closely resemble real-world observations. We demonstrate that using an 8-dimensional latent variable is sufficient for modeling catchment runoff in response to daily climate forcing variables, such as precipitation, temperature, and potential evapotranspiration, averaged over the catchment area. Here, ``latent" refers to variables not directly observed through physical properties but estimated from climate and runoff data. By sampling from this latent variable space, the generative model can create a diverse set of discharge prediction functions, each capable of generating realistic time series that resemble those observed in real-world catchment.

The generative model can be effectively applied to arbitrary catchments by estimating the optimal latent variable values using observed forcing and discharge time series, mirroring the calibration process of conventional process-based lumped hydrological models. This method was evaluated on global catchments, including those outside the generative model's training dataset. The results indicate that the model's prediction accuracy is comparable to that of conventional process-based models.

The results indicate that the hydrological behavior of catchments worldwide can be effectively characterized by a low-dimensional latent variable for daily streamflow prediction tasks. Optimal values for these latent variables can be determined solely from climate forcing and runoff time series, without the need for specific hydrological knowledge or reference to physical catchment attributes. This low-dimensional representation simplifies the modeling process and could be valuable in applications such as catchment similarity analysis and clustering in future studies.

However, when applying the generative models, we encountered challenges such as determining optimal parameter values and equifinality, which also affect conventional process-based models. Future studies should focus on methods to reliably estimate these parameters for diverse catchments. Additionally, the physical meaning of each latent variable dimension is currently unknown, and further research is encouraged to interpret these learned values.

\acknowledgments
The source code used in this research is freely available at \url{https://github.com/stsfk/deep_lumped}, where two web applications demonstrating the applications of the generative models can also be found. This study utilizes freely available datasets: Catchment Attributes and Meteorology for Large-sample Studies (CAMELS; \citeA{Newman2014_data,Newman2015,Addor2017}), Caravan \cite{Kratzert2023}, and CAMELS-DE \cite{Dolich2024,Loritz2024}. These datasets are accessible at \url{https://dx.doi.org/10.5065/D6MW2F4D}, \url{https://zenodo.org/doi/10.5281/zenodo.6522634}, and \url{https://doi.org/10.5194/essd-2024-318}, respectively. The processed CAMELS forcing data used in \citeA{Knoben2020} was provided by Wouter Knoben. The LaTex template used to create this file is from the AGU Geophysical Research Letters AGUTeX Article \url{https://www.overleaf.com/latex/templates/agu-geophysical-research-letters-agutex-article/rnyzczmyvkbj}. We thank Wouter Knoben for sharing the processed CAMELS data used in \citeA{Knoben2020} and for commenting on the paper. The results of this paper were presented at the HydroML 2024 Symposium as a tutorial and at the Information Theory in the Geosciences group, and we appreciate the feedback received at these venues. The research is supported by the Seed Fund for Basic Research for New Staff from University Research Committee (URC) of The University of Hong Kong.





%
%
%
%
%

\bibliography{agusample}

\newpage
\appendix
\section{Setting of numerical experiments}\label{appendix:raw}
\begin{table}
    \centering
\begin{tabular}{ll}
\hline
\multicolumn{2}{c}{\textbf{Settings of Bayesian optimization algorithm of hyperparameter optimization}}                                                                                                  \\ \hline
Number of trials in  & 200  \\hyperparameter optimization                                                                                                        \\
\hline
Number of  epochs    & The maximum number is 500 and stops when the error on the\\&  validation set has not improved for 20 consecutive epochs.\\
\hline
\begin{tabular}[c]{@{}l@{}}Parameters optimized \\ and possible values\end{tabular} &
  \begin{tabular}[c]{@{}l@{}}Learning rate of the embedding network: from 5e-5 to 1e-2;\\ Learning rate of the LSTM model network: 5e-5 to 1e-2;\\ Batch size: 16, 32, 64, 128, and 256;\\ LSTM layer dimension: from 4 to 256;\\ Number of LSTM layer: 1, and 2;\\ Number of fully connected layer after LSTM layer(s) for processing \\ the network output at each time step: from 1 to 3;\\ Number of neurons in the fully connected layers(s): from 2 to 32;\\ Dimension of the latent vector: 2, 4, 8, and 16;\\ Implementation of dropout: yes, and no;\\ Dropout rate: from 0.2 to 0.5 if implemented.\end{tabular} \\ \hline
\multicolumn{2}{c}{\textbf{Settings of the genetic algorithm (GA) for latent variable value optimization}}                                                                                                                \\ \hline
Population size       & 200                                                                                                                                 \\ \hline
Number of solutions to & 10 \\ be selected as parents                                                                                                                                       \\ \hline
Mutation type       & Random mutation                                                                                                                       \\ \hline
Initial population's range     & The lower and upper value of the initial population are \\ & 20\% lower or higher than the minimum and maximum value of each \\ & dimension of the optimal latent variable values of the catchments from \\& the training set.  \\ \hline
Number of generations & \begin{tabular}[c]{@{}l@{}}Optimization stop when there is no improvement for 10 consecutive\\ generations and up to 500.
\end{tabular} \\ \hline
\end{tabular}
    \caption{Settings and hyperparameter values considered in the Bayesian optimization algorithm and genetic algorithm (GA).}
    \label{table_hyperpara_used}
\end{table}

\end{document}


%
%


\title{Supporting Information for "Learning to Generate Lumped Hydrological Models"}
%
%

%
%



\authors{=Authors=}


\affiliation{=number=}{=Affiliation Address=}

%
%

%

\begin{article}

%
%

\noindent\textbf{Contents of this file}
\begin{enumerate}
\item Text S1 to Sx
\item Figures S1 to Sx
\item Tables S1 to Sx
\end{enumerate}
\noindent\textbf{Additional Supporting Information (Files uploaded separately)}
\begin{enumerate}
\item Captions for Datasets S1 to Sx
\item Captions for large Tables S1 to Sx (if larger than 1 page, upload as separate excel file)
\item Captions for Movies S1 to Sx
\item Captions for Audio S1 to Sx
\end{enumerate}

\noindent\textbf{Introduction}


\noindent\textbf{Text S1.}
%


\noindent\textbf{Data Set S1.} 


\noindent\textbf{Movie S1.} 


\noindent\textbf{Audio S1.} 
\end{article}
\clearpage